\documentclass[journal]{IEEEtran}
\usepackage{ifpdf}

\usepackage{cite}

\ifCLASSINFOpdf
  \usepackage[pdftex]{graphicx}
  \graphicspath{{figs/}{}}
  \DeclareGraphicsExtensions{.pdf,.jpeg,.png}
  \usepackage{epstopdf}
\else
  
  \usepackage[dvips]{graphicx}
  \graphicspath{{figs/}}
  \DeclareGraphicsExtensions{.eps}
\fi
\usepackage[cmex10]{amsmath}
\usepackage{array}

\usepackage[normalem]{ulem}	

\usepackage{verbatim}

\usepackage{color}

\usepackage{version}
\usepackage{siunitx}

\excludeversion{notes}

\begin{document}
%
\title{Permittivity and Conductivity Measured using a Novel Toroidal Split-Ring Resonator}
%
%
%


\author{Jake~S.~Bobowski 
       and~Aaron~P.~Clements
\thanks{Manuscript submitted on \today.}
\thanks{J.S.~Bobowski and A.P.~Clements are with the Department
of Physics, University of British Columbia, Kelowna,
BC, V1V 1V7 Canada (e-mail: \mbox{Jake.Bobowski@ubc.ca}).}
}

%
%


\markboth{Bobowski \MakeLowercase{\textit{et al.}}: Permittivity and Conductivity Measured using a Novel Toroidal Split-Ring Resonator}{}
%



\maketitle

\begin{abstract}
We describe and demonstrate the use of a novel toroidal split-ring resonator to make accurate and precise measurements of the permittivity of liquids and gases.  We first analytically show how the resonance frequency and quality factor of the resonator are related to the complex permittivity of the material filling its gap.  We then use the resonator to experimentally determine the permittivity of a number of different materials. First, the compact and high-\textbf{\textit{Q}} resonator is used to measure both the real and imaginary parts of the complex permittivity of methyl alcohol at 185~MHz.  Second, the resonator was placed inside a vacuum-tight Dewar.  We measured the  resonance frequency with the resonator suspended in vacuum and then immersed in an atmosphere of air.  From these data, the dielectric constant of air was accurately determined.  Next, the resonator was submerged in liquid nitrogen and the boiling temperature of the nitrogen bath was manipulated by regulating its vapor pressure.  This system allowed for a precise measurement of the dielectric constant of liquid nitrogen over a temperature range of 64 to 77~K.  Finally, we monitored the quality factor of the copper resonator as its temperature drifted from 80~K to room temperature.  From these data, we extracted the linear temperature dependence of copper's resistivity.
\end{abstract}


%
\IEEEpeerreviewmaketitle

\section{Introduction}
%
%
%
%

\IEEEPARstart{T}{he} resonant cavity techniques developed six decades ago for electromagnetic (EM) material property measurements continue to be used to this day \cite{Sproull:1946, Birnbaum:1949,Peng:2014, Orloff:2014, Han:2015}.  The simplest cavity resonators are relatively easy to build and operate and, because of their high quality factors, provide results that are both precise and accurate \cite{Krupka:2006}.  However, determining the EM properties of a material is only straightforward when the sample under test matches the symmetry of the cavity.  Furthermore, because of their large size, cavity resonators are rarely used below \SI{2}{\giga\hertz}.  At these frequencies it is impractical to achieve the filling factors needed to maintain the measurement sensitivity \cite{Krupka:2006}.

For permittivity measurements in the range of \SI{50}{\mega\hertz} to \SI{2}{\giga\hertz}, reentrant cavities can be used \cite{Krupka:2006}. However, in this case, the data analysis requires numerical methods \cite{Kaczkowski:1980A, Kaczkowski:1980B}.  In 1981, Hardy and Whitehead developed the split-ring resonator (SRR), also known as a loop-gap resonator which, like re-entrant cavities, has a useful frequency range from tens of megahertz to several gigahertz \cite{Hardy:1981}.  Since their development, SRRs have been used in novel ESR (electron spin resonance) spectrometers \cite{Froncisz:1982} and, more recently, as components of negative refractive index metamaterials \cite{Smith:2000}.  In this paper, we describe and demonstrate how a novel toroidal SRR can be used to make high-resolution measurements of the complex permittivity of various materials, while using only standard data analysis methods.

Cylindrical split-ring resonators (SRRs) consist of a conducting tube with a narrow slit along its length. Magnetic flux can be coupled into and out of the resonator using coupling loops placed at either end of the resonator.  The bore of the resonator has an intrinsic inductance $L_0$, the slit or gap has capacitance $C_0$, and the effective resistance $R_0$ is determined by the resistivity and skin depth of the conductor.  As a result, the resonator can be modelled as an $LRC$ circuit with corresponding vacuum resonance frequency \mbox{$\omega_0=2\pi f_0\approx 1/\sqrt{L_0 C_0}$} and quality factor \mbox{$Q_0\approx R_0^{-1}\sqrt{L_0/C_0}$} \cite{Bobowski:2013}.  

Placing a dielectric material in the gap of the resonator will result in a change to the effective capacitance. In a similar manner, a magnetic material in the bore will modify the inductance. Thus, the presence of these materials will cause both the resonance frequency and quality factor of the SRR to deviate from their intrinsic, or vacuum, values.  These deviations can be used to determine the real and imaginary components of the materials' complex permittivity and/or permeability \cite{Bobowski:2015}.  For example, in previous work we used cylindrical SRRs  to measure the dielectric constant of distilled water and the conductivity of various concentrations of NaCl dissolved in water \cite{Bobowski:2013}.  In another remarkable application, ultra-high-$Q$ cylindrical SRRs have been fabricated by electroplating the copper resonator with an alloy (lead, 5\% tin) that becomes superconducting when cooled using liquid $^4$He \cite{Bonn:1991}.  These devices have been used to measure temperature dependence of the magnetic penetration depth of the high-temperature superconductor YBa$_2$Cu$_3$O$_{6.95}$ \cite{Hardy:1993}. 

A disadvantage of the cylindrical geometry is that, on resonance, the currents induced on the inner wall of the resonator generate magnetic field lines that extend outside the bore of the resonator and into free space.  These radiative losses can substantially degrade the $Q$ of the resonance.  Additional EM shielding can be used to suppress the radiative losses; however, these shields can be bulky and inconvenient in many applications \cite{Hardy:1981, Bobowski:2013}. 

\begin{figure}[t]
\centering{(a)\quad\includegraphics[keepaspectratio, width=0.7\columnwidth]{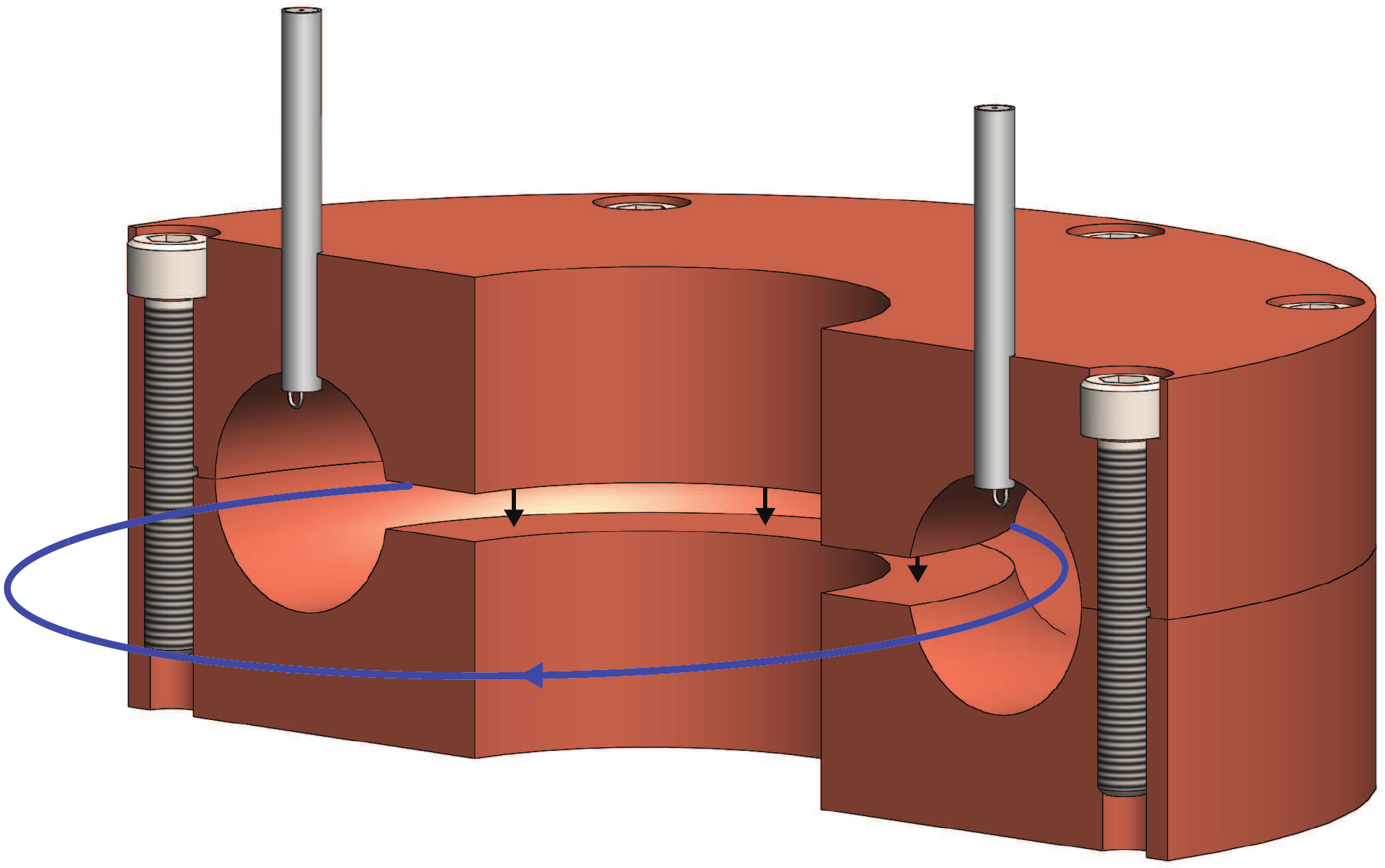}\\(b)\quad\includegraphics[keepaspectratio, width=0.7\columnwidth]{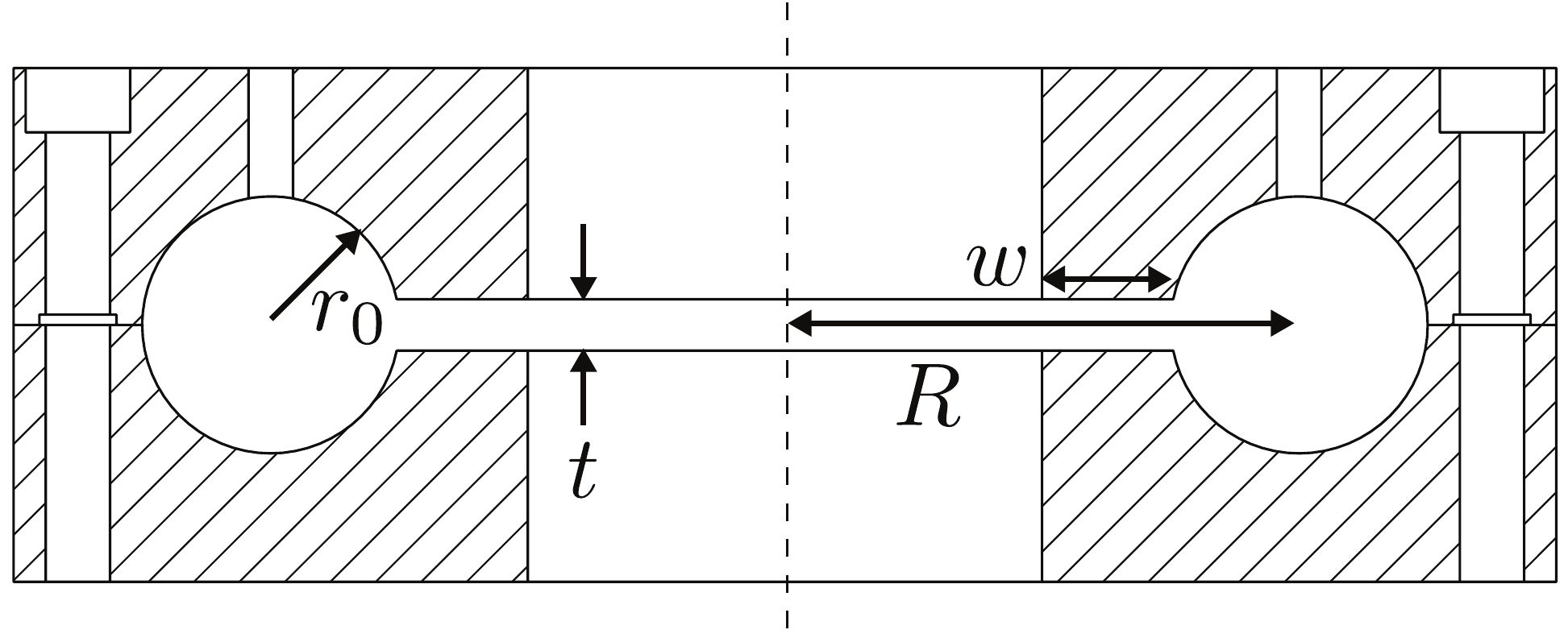}}
\caption{\label{fig:TSRR}(a) A cutaway view of the toroidal SRR with coupling loops in place.  Representative electric field lines (black) in the gap of the resonator and magnetic field lines (blue) in the bore of the resonator are shown. (b) A scale drawing of the toroidal SRR cross-section. For clarity, the size of the gap has been exaggerated.  The dimensions of the toroidal SRRs used in this work are $r_0=w=\SI{0.635}{\centi\meter}$ (\SI{0.250}{in}.) and $R=\SI{2.54}{\centi\meter}$ (\SI{1.00}{in}.).  The measurements in Section~\ref{sec:gas} were made using a SRR with a gap dimension \mbox{$t=\SI{0.889}{\milli\meter}$}, all other measurements used a SRR with \mbox{$t=\SI{0.254}{\milli\meter}$}.}
\end{figure}

Recently, we described the design and characterization of a novel toroidal SRR \cite{Bobowski:2016}.  As shown in Fig.~\ref{fig:TSRR}, in the toroidal geometry magnetic field lines are completely contained within the bore of the resonator, such that high intrinsic quality factors are obtained without requiring additional shielding.  This design offers a stable and compact resonator that is ideally suited for the characterization of the EM properties of liquids and gases.  

In Section~\ref{sec:theory} of this paper, we show how the resonance frequency and $Q$ of the SRR are related to the complex permittivity of the material filling its gap.   We then describe several applications of the toroidal SRR.  Section~\ref{sec:meth} describes  a measurement of the complex permittivity of methyl alcohol (methanol) at \SI{185}{MHz}.  Section~\ref{sec:gas} demonstrates the sensitivity of the toroidal SRR by using it to measure the dielectric constant of the atmosphere of air.  In Section~\ref{sec:LN2}, the toroidal SRR is used to measure the dielectric constant of liquid nitrogen as a function of its temperature.  Finally, in Section~\ref{sec:CuSigma}, we use the temperature dependence of the resonator's quality factor to extract the linear temperature dependence of copper's resistivity.  Section~\ref{sec:summary} summarizes the results and briefly discusses some possible future applications of the toroidal SRR.

\section{Theory}\label{sec:theory}
The signal detected by the output coupling loop is proportional to the magnetic flux in the bore of the resonator which is induced by currents that run along the inner wall of the SRR.  The magnitude and phase of the current is, in turn, determined by the total effective impedance of the resonator.  As described in \cite{Bobowski:2013}, the effective resistance $R$ of the SRR is inversely proportional to the EM skin depth $\delta$~\cite{Hardy:1981, Bobowski:2016}.  For a good conductor, \mbox{$\delta\propto 1/\sqrt{\omega}$}, such that
\begin{equation}
R=R_0\sqrt{\frac{\omega}{\omega_0}}
\end{equation}
where $R_0$ is the resistance at the vacuum resonance frequency $\omega_0$.

The resonant properties of the SRR can be modified by submerging the resonator in a fluid with relative permittivity \mbox{$\varepsilon_\mathrm{r}=\varepsilon^\prime-j\varepsilon^{\prime\prime}$}.  In this work, we have adopted the wave propagation sign convention $e^{j\omega t}$ such that $\varepsilon^{\prime\prime}$ is positive.  The fluid within the gap of the SRR, where the electric field is concentrated, will result in an enhanced effective capacitance \mbox{$C_1=\varepsilon_\mathrm{r}C_0$}.  If the fluid is non-magnetic, the inductance is unaffected by its presence and the effective SRR impedance becomes
\begin{align}
&Z_1(\omega)=R_0\sqrt{\frac{\omega}{\omega_0}}+j\omega L_0+\frac{1}{j\omega\varepsilon_\mathrm{r} C_0}\\
&=\left[R_0\sqrt{\frac{\omega}{\omega_0}}+\frac{\varepsilon^{\prime\prime}}{\left\vert\varepsilon_\mathrm{r}\right\vert^2}\frac{1}{\omega C_0}\right]+j\left[\omega L_0-\frac{\varepsilon^\prime}{\left\vert\varepsilon_\mathrm{r}\right\vert^2}\frac{1}{\omega C_0}\right]
\end{align}
where \mbox{$\left\vert\varepsilon_\mathrm{r}\right\vert^2\equiv\left(\varepsilon^\prime\right)^2+\left(\varepsilon^{\prime\prime}\right)^2$}.  The imaginary component of $\varepsilon_\mathrm{r}$ results in a second contribution to the real part of $Z_1$ and, therefore, additional losses.  Expressed in terms of $\omega_0$ and $Q_0$, the magnitude and phase of $Z_1^{-1}$ are given by
\begin{align}
\frac{R_0 Q_0}{\left\vert Z_1(\omega)\right\vert}&=\left\{\left[\frac{1}{Q_0}\sqrt{\frac{\omega}{\omega_0}}+\frac{\varepsilon^{\prime\prime}}{\left\vert\varepsilon_\mathrm{r}\right\vert^2}\frac{\omega_0}{\omega}\right]^2\right.\nonumber\\
&\qquad\qquad\qquad\left.+\left[\frac{\omega}{\omega_0}-\frac{\varepsilon^\prime}{\left\vert\varepsilon_\mathrm{r}\right\vert^2}\frac{\omega_0}{\omega}\right]^2\right\}^{-1/2}\label{eq:Z1}\\
\tan\phi&=\frac{\dfrac{\varepsilon^\prime}{\left\vert\varepsilon_\mathrm{r}\right\vert^2}\dfrac{\omega_0}{\omega}-\dfrac{\omega}{\omega_0}}{\dfrac{1}{Q_0}\sqrt{\dfrac{\omega}{\omega_0}}+\dfrac{\varepsilon^{\prime\prime}}{\left\vert\varepsilon_\mathrm{r}\right\vert^2}\dfrac{\omega_0}{\omega}}.\label{eq:phi}
\end{align}
If $\omega_0$ and $Q_0$ are known, the SRR response can be fit to the frequency dependence of $\left\vert Z_1(\omega)\right\vert^{-1}$ and $\tan\phi$ to determine experimental values of $\varepsilon^\prime$ and $\varepsilon^{\prime\prime}$.

When the $\varepsilon^{\prime\prime}$ term of $Z_1$ is sufficiently small and $\varepsilon_\mathrm{r}(\omega)$ remains approximately constant over the bandwidth of the resonance, $\left\vert Z_1\left(\omega\right)\right\vert^{-1}$ will be a sharply peaked function of frequency and its shape will remain nearly Lorentzian.  Under these conditions, the modified resonance frequency $\omega_1$ and quality factor $Q_1$ of the SRR will be given by
\begin{eqnarray}
\omega_1 &\approx &\frac{\sqrt{\varepsilon^\prime(\omega_1)}}{\left\vert\varepsilon_\mathrm{r}(\omega_1)\right\vert}\omega_0\label{eq:w1}\\
\frac{1}{Q_1} &\approx & \frac{1}{Q_0}\left[\frac{\left\vert\varepsilon_\mathrm{r}(\omega_1)\right\vert^2}{\varepsilon^\prime(\omega_1)}\right]^{1/4}+\frac{\varepsilon^{\prime\prime}(\omega_1)}{\varepsilon^\prime(\omega_1)}.\label{eq:Q1}
\end{eqnarray}
These approximate expressions are valid when both $Q_0$ and $Q_1$ are much greater than one.  Solving (\ref{eq:w1}) and (\ref{eq:Q1}) for the real and imaginary components of the permittivity results in
\begin{align}
\varepsilon^\prime &\approx\dfrac{\left(\dfrac{\omega_0}{\omega_1}\right)^2}{1+\left[\dfrac{1}{Q_1}-\dfrac{1}{Q_0}\left(\dfrac{\omega_0}{\omega_1}\right)^{1/2}\right]^2}\label{eq:eprime}\\
\varepsilon^{\prime\prime} &\approx\dfrac{\left(\dfrac{\omega_0}{\omega_1}\right)^2 \left[\dfrac{1}{Q_1}-\dfrac{1}{Q_0}\left(\dfrac{\omega_0}{\omega_1}\right)^{1/2}\right]}{1+\left[\dfrac{1}{Q_1}-\dfrac{1}{Q_0}\left(\dfrac{\omega_0}{\omega_1}\right)^{1/2}\right]^2}\label{eq:edprime}
\end{align}
where the subscript ``0'' refers to values that would be obtained from measurements with the resonator suspended in vacuum and those with subscript ``1'' refer to values with the SRR immersed in the fluid~\cite{Bobowski:2015}.

\section{Experimental Measurements}
\subsection{Methanol permittivity}\label{sec:meth}
\begin{figure}[t]
\centering{\includegraphics[width=\columnwidth]{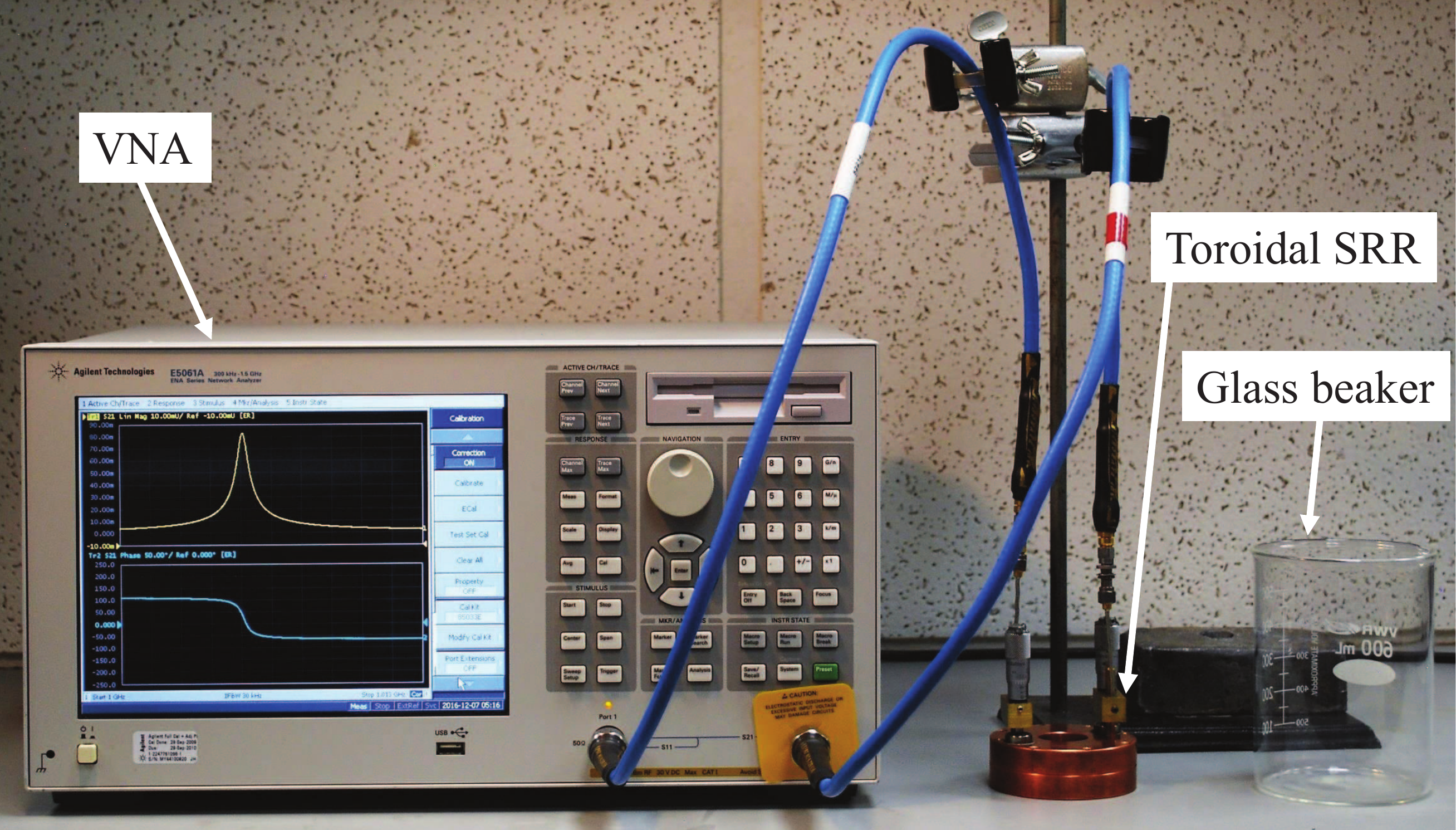}}
\caption{\label{fig:methExpt}Digital photograph of the experimental set up used to collect the data shown in Section~\ref{sec:meth}.  The coupling loops are positioned using translation stages controlled by micrometers.  Also shown is the glass beaker that was used to submerge the resonator in methanol.}
\end{figure}

In the measurements presented in this section, the resonator's input coupling loop was driven by port~1 of an Agilent N5241A vector network analyzer (VNA) and the power coupled out of the resonator was detected by port~2.  An Agilent N4691-60004 Electronic Calibration (ECal) module was used to establish a calibration plane at the free ends of the low-loss and high-stability cables connected to ports 1 and 2 of the VNA.  The pair of coupling loops were made using short ($\sim\SI{10}{\centi\meter}$) sections of \SI{2.16}{\milli\meter}-diameter (\SI{0.085}{in}.) semi-rigid coaxial cables with a copper outer conductors and a silver-plated copper-weld (SPCW) center conductors. For the purpose of this work, the semi-rigid cables used for the coupling loops were assumed to be lossless.  All measurements reported in this section were obtained by averaging 1000 VNA frequency sweeps.  In this work, no attempt has been made to account for residual errors associated with VNA ECal system.  The experimental set up is shown in Fig.~\ref{fig:methExpt}.

\begin{figure}[t]
\centering{\includegraphics[width=0.95\columnwidth]{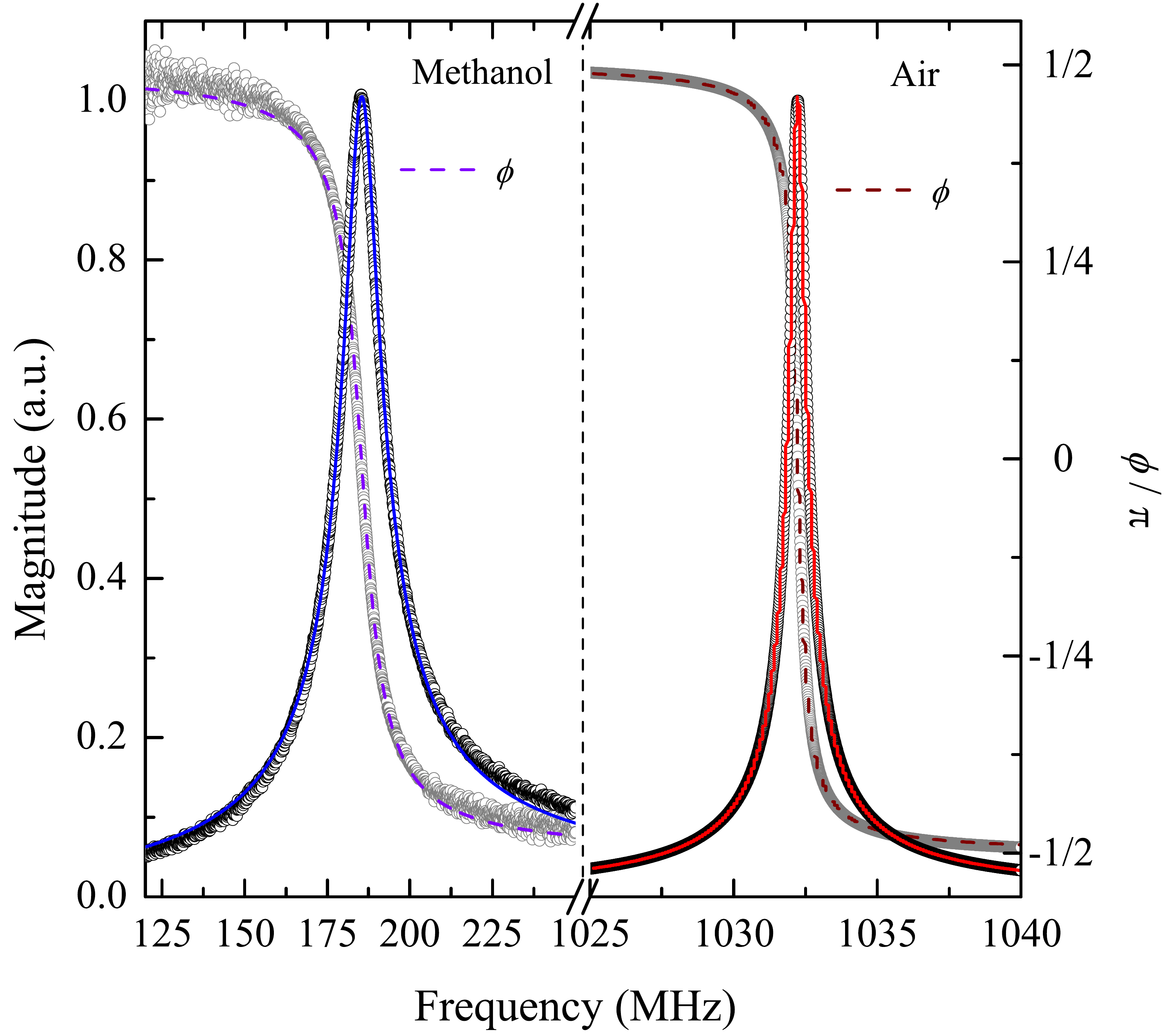}}
\caption{\label{fig:meth}Magnitude and phase of the toroidal SRR resonance. Left: When the SRR is submerged in methyl alcohol. Right: When the SRR is surrounded by air.  The magnitude data have been scaled to be equal to one at the resonance frequencies.  The solid lines are fits to (\ref{eq:Z1}) and the dashed lines are fits to (\ref{eq:phi}).}
\end{figure}

First, the \SI{7.6}{\centi\meter}-diameter (\SI{3}{in}.) toroidal SRR described in \cite{Bobowski:2016} was assembled and the magnitude and phase of the resonance were measured while the resonator was surrounded by air.  The results are shown on the right-hand side of Fig.~\ref{fig:meth}. 
The $\phi$-data were corrected for the additional phase shift introduced by the cables used to create the two coupling loops \cite{Bobowski:2016}.  As will be discussed in the next section, the dielectric constant of \SI{1}{atm} of air is $\varepsilon^\prime_\mathrm{air} = 1.00059$.  Using MATLAB\textsuperscript{\textregistered}, the air magnitude and phase data were simultaneously fitted to (\ref{eq:Z1}) and (\ref{eq:phi}) to determine $\omega_0$ and $Q_0$.  For these fits, we set $\varepsilon^{\prime\prime}$ to zero and used \mbox{$\varepsilon^\prime/\left\vert\varepsilon_\mathrm{r}\right\vert = 1/\varepsilon^\prime_\mathrm{air}$}.  The best-fit values for the vacuum resonance frequency and quality factor are summarized in Table~\ref{tab:fits}.
\begin{table}[t]
\caption{\label{tab:fits}Summary of fit results to the data in Fig.~\ref{fig:meth} (\mbox{$\Delta f_0 =\SI{400}{Hz}$}).}
\centering
\begin{tabular}{cc}
\\[-1ex]
$f_0= \omega_0/2\pi $ (\SI{}{MHz}) & $Q_0$\\[1.1ex]
\hline\hline\\[-1ex]
~\quad$1032.52$\quad~ & ~\quad$2024\pm 3$\quad~\\[1.1ex]
\hline\\[1.5ex]
$\varepsilon^\prime$ & $\varepsilon^{\prime\prime}$\\[1.1ex]
\hline\hline\\[-1ex]
~\quad$30.978\pm 0.006$\quad~ & ~\quad$1.703\pm 0.006$\quad~\\[1.1ex]
\hline
\end{tabular}
\end{table}

Next, the toroidal SRR was placed into a glass beaker and submerged in 99\% pure methyl alcohol at room temperature.  Care was taken to ensure that the liquid completely filled the bore and gap of the resonator.  Changes to either the electrical contact between the two halves of the SRR or the positions of the coupling loops can affect both resonance frequency and quality factor \cite{Bobowski:2016}.  Therefore, it is crucial that these properties of the resonator remain unchanged between the air and methanol measurements.  The methanol results are shown on the left-hand side of Fig.~\ref{fig:meth}. Uncertainties were estimated using the observed scatter in the datasets.  For the magnitude measurements, the uncertainty is approximately independent of frequency while the phase uncertainty is a minimum at the resonance frequency and increases as $\phi$ approaches $\pm\pi/2$.  

Simultaneous weighted fits to (\ref{eq:Z1}) and (\ref{eq:phi}) were once again performed.  However, this time the known values of $\omega_0$ and $Q_0$ from the air analysis were used and $\varepsilon^\prime$ and $\varepsilon^{\prime\prime}$ were the fit parameters.  The best fit values for the complex permittivity of methanol at \SI{185.2}{MHz} are summarized in Table~\ref{tab:fits}.

The uncertainties in $f_0$, $Q_0$, $\varepsilon^\prime$, and $\varepsilon^{\prime\prime}$ were determined using the 95\% confidence intervals from MATLAB's nonlinear fit routine.  Through repeated measurements, we experimentally verified that the $\Delta f_0$ and $\Delta Q_0$ estimates correctly characterize the stability of the air-filled resonator. We also verified that the extracted $\varepsilon^\prime\pm\Delta\varepsilon^\prime$ and $\varepsilon^{\prime\prime}\pm\Delta\varepsilon^{\prime\prime}$ remain unchanged when the values of $f_0$ and $Q_0$ used for the fitting were varied by $\Delta f_0$ and $\Delta Q_0$, respectively.

Our room-temperature \mbox{($21\pm 1^\circ$C)} measurement of methanol's relative permittivity is in reasonably good agreement with the broadband measurements by Bao {\it et al.\/} at $28^\circ$C who report \mbox{$\varepsilon_\mathrm{r}=\left(33.1\pm 0.8\right) + j \left(1.63\pm 0.07\right)$} at \SI{185.2}{MHz} \cite{Bao:1996}.  However, as is typical with resonant methods, the measurements made using the toroidal SRR offer far more precision.  The obvious disadvantage of the toroidal SRR measurement is that one gets a measurement of the permittivity at only a single frequency.  At the conclusion of this paper, we briefly discuss a variation of the toroidal SRR measurement that may allow for high-precision spectroscopy. 

\subsection{Dielectric constant of air}\label{sec:gas}
\begin{figure}[t]
\centering{\includegraphics[width=\columnwidth]{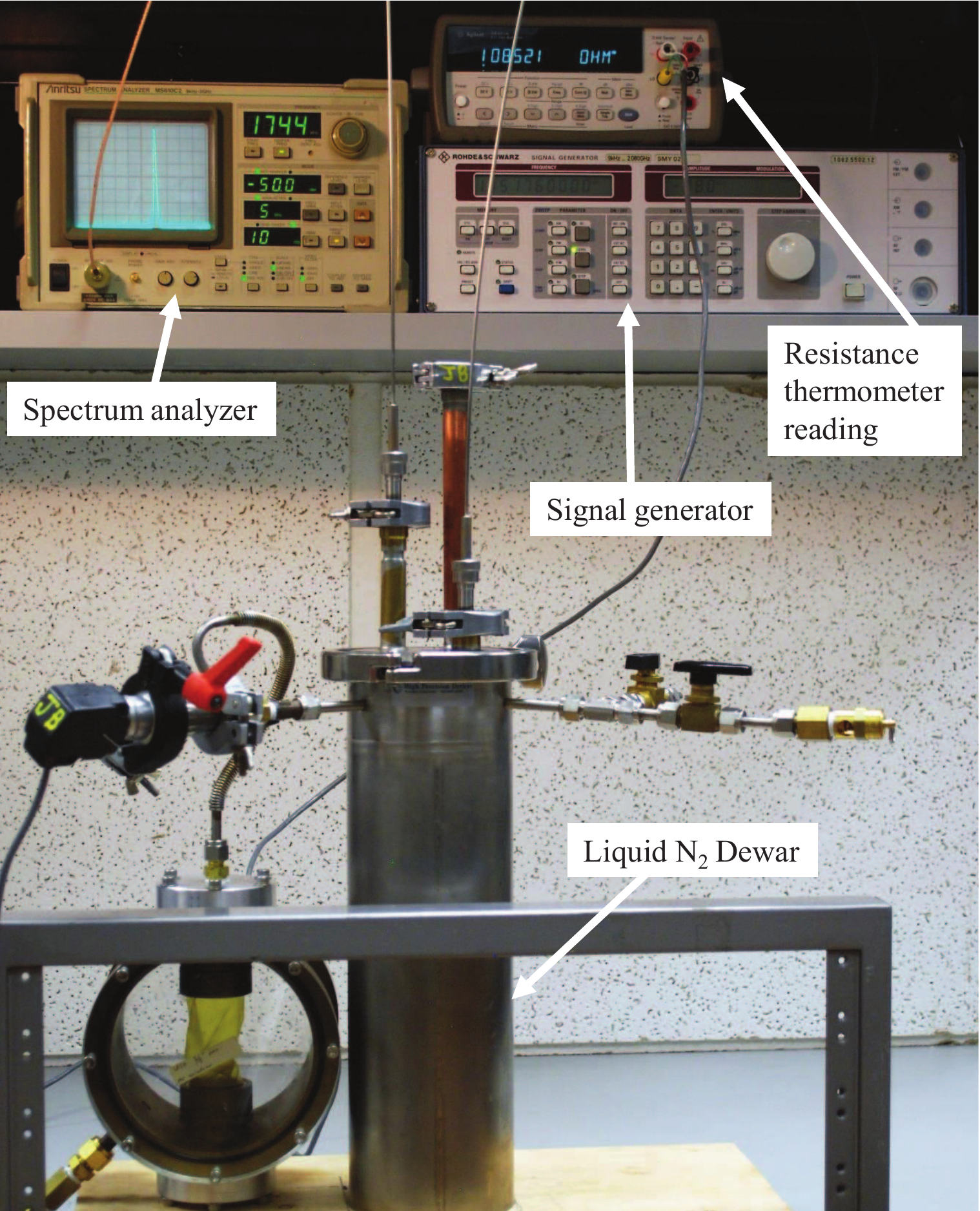}}
\caption{\label{fig:LN2Expt}Digital photograph of the Dewar and test equipment used to measure the dielectric constants of air and liquid nitrogen and to characterize the temperature dependence of the resistivity of the copper.}
\end{figure}
\begin{figure}[t]
\centering{\includegraphics[width=0.85\columnwidth]{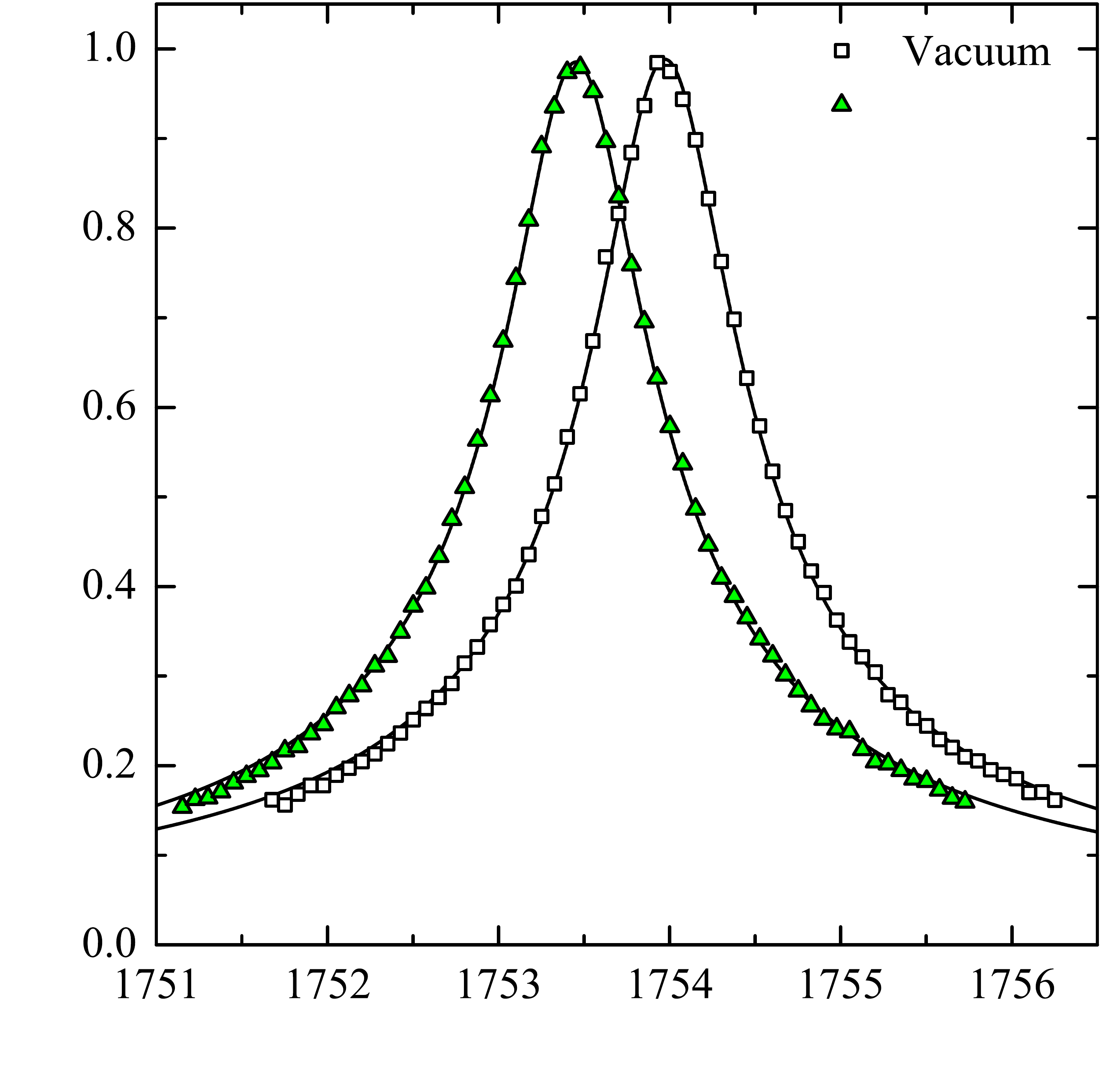}}
\caption{\label{fig:N2gas}Measured resonance of the toroidal SRR when it is suspended in vacuum (white squares) and immersed in an atmosphere of air (green triangles).  The magnitude data have been scaled to be equal to one at the resonance frequencies.  The solid lines are fits to (\ref{eq:Z1}).}
\end{figure}

For the remainder of the measurements reported in this paper, the toroidal SRR was placed into a vacuum-tight stainless steel Dewar.  The resonator's input coupling loop was driven using a Rohde \& Schwarz SMY~02 signal generator and the signal coupled out was detected using an Anritsu MS610C2 spectrum analyzer.  The experimental set up is shown in Fig.~\ref{fig:LN2Expt}. We also note that the measurements in this section were taken using a toroidal SRR with a gap size of $t=\SI{0.889}{\milli\meter}$.  All other reported measurements were made using a resonator that had a gap size of \SI{0.254}{\milli\meter}.  Other than the gap size, the two resonators were identical and as described in \cite{Bobowski:2016}.  The larger gap size reduces the effective capacitance of the resonator thereby increasing its resonance frequency. (Compare the right-hand frame of Fig.~\ref{fig:meth} to the data in Fig.~\ref{fig:N2gas}.)

First, the Dewar was evacuated using an Alcatel direct-drive mechanical pump.  A stable base pressure of \SI{350}{\milli Torr} was achieved (0.05\% of an atmosphere) and the SRR resonance was measured.  The data are shown as the white squares in Fig.~\ref{fig:N2gas}.  Next, the pump was turned off, the Dewar was filled with room-temperature air, and the resonance remeasured (green triangles in Fig.~\ref{fig:N2gas}).

First, the vacuum data were fitted to (\ref{eq:Z1}) with $\varepsilon^\prime=1$ and $\varepsilon^{\prime\prime}=0$ to determine $\omega_0$ and $Q_0$.  The fit is shown in the figure and the extracted parameters are given in Table~\ref{tab:gases}.  The air data were then fitted to (\ref{eq:Z1}) using the extracted $\omega_0$ and $Q_0$ values.
The loss tangent \mbox{$\left(\tan\delta\equiv\varepsilon^{\prime\prime}/\varepsilon^\prime\right)$} of air is very small at RF frequencies such that $\varepsilon^{\prime\prime}\approx 0$.  In this case, the only fit parameters are an overall scaling factor set by $R_0$ and $\varepsilon^\prime_\mathrm{air}$.  Our value for the dielectric constant of air, given in Table~\ref{tab:gases}, is reasonably close to that of Hector and Schultz who report \SI{1.00058986(43)} at NPT (\SI{1}{atm} and $20^\circ$C) and \SI{900}{kHz} \cite{Hector:1936}.  The uncertainty in $\varepsilon^\prime_\mathrm{air}$ was again determined using the 95\% confidence interval of the nonlinear fit routine.
\begin{table}[t]
\caption{\label{tab:gases}Summary of fit results to the data in Fig.~\ref{fig:N2gas} (\mbox{$\Delta f_0 =\SI{2}{\kilo\hertz}$}).}
\centering
\begin{tabular}{cc}
\\[-1ex]
$f_0= \omega_0/2\pi $ (\SI{}{MHz}) & $Q_0$\\[1.1ex]
\hline\hline\\[-1ex]
~\quad$1753.97$\quad~ & ~\quad$2234\pm 13$\quad~\\[1.1ex]
\hline\\[1.5ex]
\multicolumn{2}{c}{$\varepsilon^\prime$}\\[1.1ex]
\hline\hline\\[-1ex]
\multicolumn{2}{c}{~\quad$\SI{1.0005940(21)}{}$\quad~}\\[1.1ex]
\hline
\end{tabular}
\end{table}

Before leaving this section we note that, since the vacuum and air resonances are very sharply peaked, both datasets in Fig.~\ref{fig:N2gas} are Lorentzian to a very good approximation such that
\begin{equation}
\frac{R_{0,1} Q_{0,1}}{\left\vert Z_{0,1}(\omega)\right\vert}\approx\frac{1}{\sqrt{\dfrac{1}{Q_{0,1}^2}+\left(\dfrac{\omega}{\omega_{0,1}}-\dfrac{\omega_{0,1}}{\omega}\right)^2}}\label{eq:Lorentz}
\end{equation} 
where the ``0'' subscripts represents vacuum and the ``1'' is for air.  In this case, both $Q_0$ and $Q_1$ are much larger than one and, from (\ref{eq:eprime}), the real part of the permittivity, or dielectric constant of air, is given by \mbox{$\varepsilon_\mathrm{r}\approx\varepsilon^\prime_\mathrm{air}\approx\left(\omega_0/\omega_1\right)^2$}.  Fitting the data in Fig.~\ref{fig:N2gas} to (\ref{eq:Lorentz}) yields \mbox{$f_0=\SI{1753.9732(13)}{MHz}$} and \mbox{$f_1=\SI{1753.4525(10)}{MHz}$}.  This simplified analysis gives \mbox{$\varepsilon^\prime_\mathrm{air}=\SI{1.0005939(19)}{}$} which is in very good agreement with the result obtained from the full analysis that is quoted in Table~\ref{tab:gases}.

\subsection{Dielectric constant of liquid nitrogen}\label{sec:LN2}
For the next set of measurements, the SRR was mounted on a Teflon stand, so as to thermally isolate it from the walls of the Dewar, and then submerged in liquid nitrogen.  The N$_2$ vapor above the liquid was connected to a \SI{15}{\litre} buffer volume whose pressure was regulated using a needle valve and the Alcatel mechanical pump.  Pumping a large volume of gas through a large impedance (set by the needle valve) allows one to very slowly and controllably reduce the vapor pressure and, therefore, the boiling temperature of the nitrogen bath.  A calibrated platinum resistance thermometer from Cryogenic Control Systems, Inc.\ was attached to the toroidal SRR which allowed us to characterize the resonance as a function of the temperature of the liquid N$_2$ bath.  Once again, as the measurements of Reesor {\it et al.\/} show, the loss tangent of liquid N$_2$ is very small at RF frequencies such that \mbox{$\varepsilon_\mathrm{r}\approx\varepsilon^\prime_{\mathrm{N}_2}\approx\left(\omega_0/\omega_1\right)^2$} \cite{Reesor:1975}.
\begin{figure}[t]
\centering{\includegraphics[width=0.9\columnwidth]{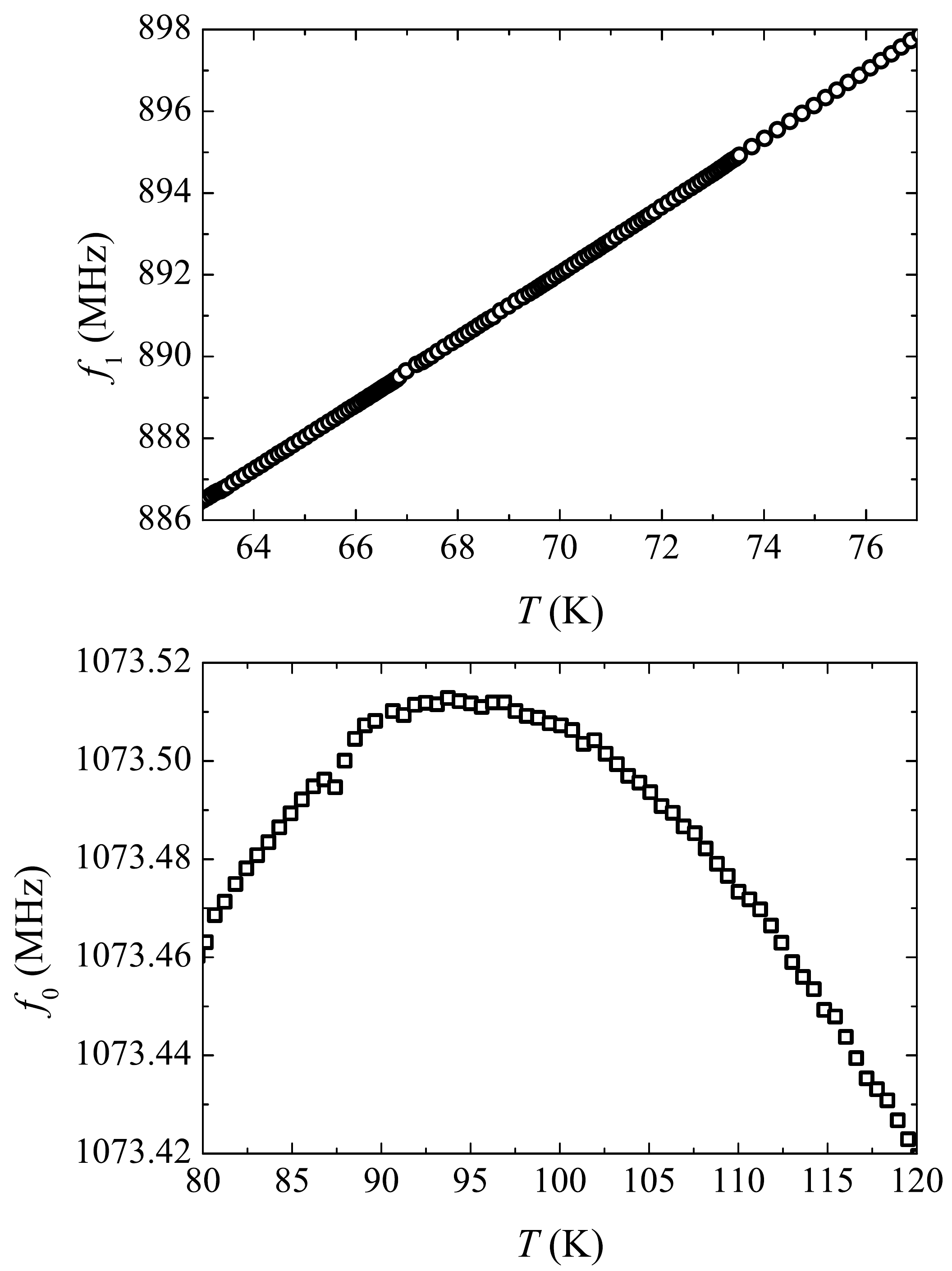}}
\caption{\label{fig:drift}Top: The toroidal SRR resonance frequency $f_1$ when it is submerged in liquid nitrogen as a function of temperature.  Bottom: The resonance frequency $f_0$ versus temperature when the level of the nitrogen bath is below the resonator.}
\end{figure}

Figure~\ref{fig:drift} shows how resonance frequency of the liquid-filled resonator changed as the temperature of the bath was varied between its freezing temperature (\SI{63}{K}) and its boiling temperature at \SI{1}{atm} (\SI{77}{K}).  Over this temperature range, $f_1$ varied linearly with a slope of \SI{0.81}{MHz\per K}.  We also monitored the temperature dependence of the resonance frequency when there was no liquid in the resonator.  The data from \SI{80} to \SI{120}{K} are shown in the bottom half of Fig.~\ref{fig:drift}.  In this case, the variation in $f_0$ is due to a combination of two effects.  First, the nitrogen vapor above the liquid modifies the capacitance of the SRR.  Second, the thermal expansion of the copper will result in small changes to both the capacitance and inductance of the resonator.  However, in the temperature range of interest, the maximum observed slope of $f_0$ versus $T$ was \SI{0.005}{MHz\per K} which is more than 150 times smaller than $df_1/dT$ and can, therefore, be neglected.   

\begin{figure}[t]
\centering{\includegraphics[width=0.95\columnwidth]{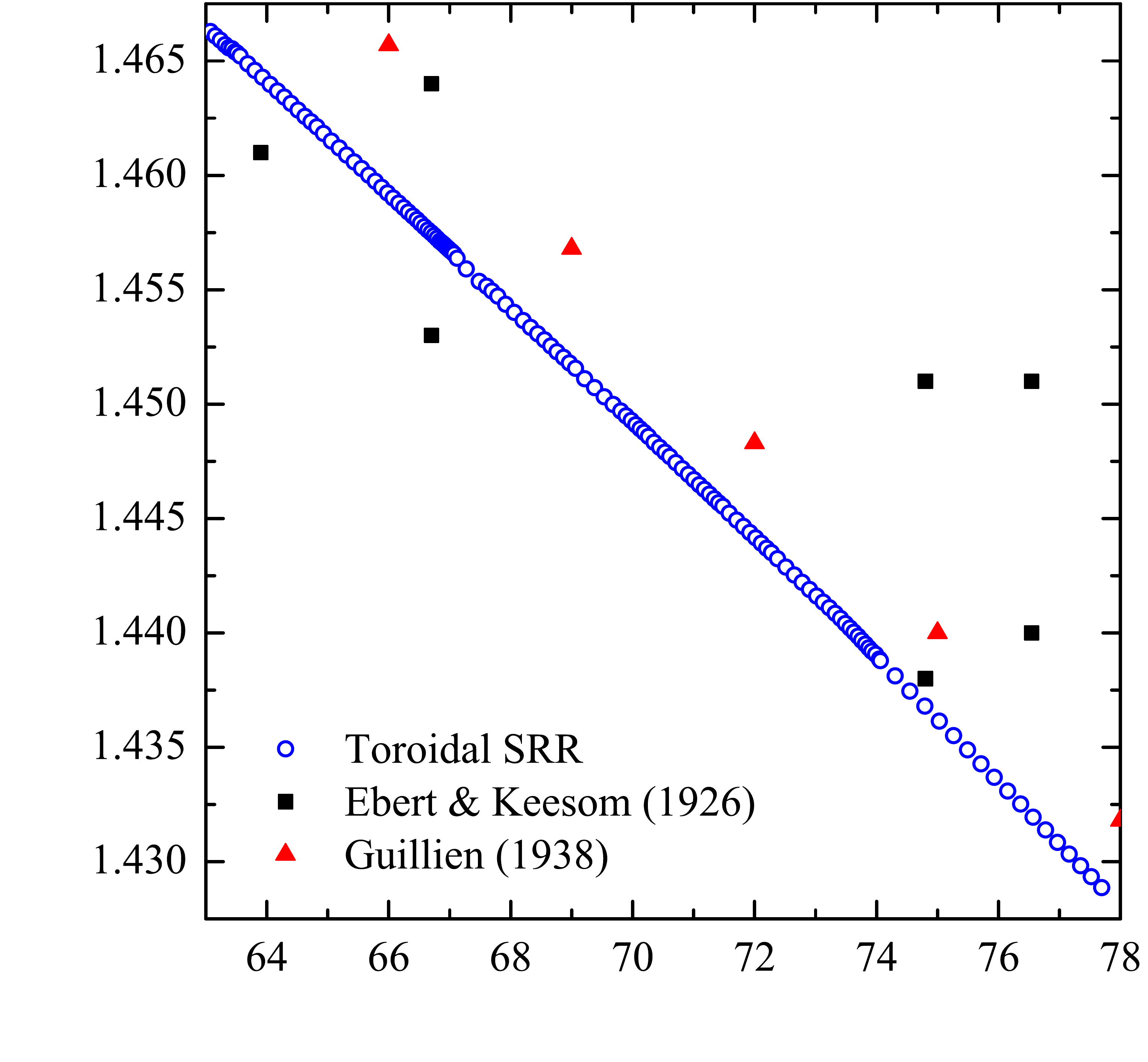}}
\caption{\label{fig:nitrogen}The dielectric constant of liquid nitrogen as a function of temperature.  The high-resolution toroidal SRR data are shown using blue circles.  Early measurements of liquid nitrogen's dielectric constant obtained from \cite{Jensen:1980} are shown using squares and triangles.}
\end{figure}

Using Fig.~\ref{fig:drift} to set $f_0 = \SI{1073.46}{MHz}$, the $f_1$ measurement can be used to extract the temperature dependence of the dielectric constant of liquid nitrogen.  The results are shown in Fig.~\ref{fig:nitrogen} using blue circles.  
The uncertainty in the $\varepsilon^\prime_{\mathrm{N}_2}$ measurements is dominated by the drift in the liquid N$_2$ bath temperature during each frequency sweep.  The $f_1$ data shown in Fig.~\ref{fig:drift} were collected over a period of 70 minutes which corresponds to an average $f_1$ drift rate of less than \SI[per-mode=symbol]{0.2}{\mega\hertz\per\minute}.  Each frequency sweep took \SI{30}{\second} to complete which gives $\Delta f_1 \approx \SI{0.1}{\mega\hertz}$.  Therefore, \mbox{$\Delta\varepsilon^\prime_{\mathrm{N}_2}=\left(f_0/f_1\right)^2\Delta f_1/f_1\approx 1\times 10^{-4}$} which is smaller than the size of the data points in Fig.~\ref{fig:nitrogen}.

For comparison, Fig.~\ref{fig:nitrogen} also includes early measurements of liquid nitrogen's dielectric constant obtained from \cite{Jensen:1980} (squares and triangles). We emphasize that the toroidal SRR measurements are relatively easy and quick to make and the data analysis is very simple.  Most importantly, the high-precision measurements of $\varepsilon^\prime_{\mathrm{N}_2}$ are accurate and allow for a finely-spaced dataset.

\subsection{Resistivity of Copper}\label{sec:CuSigma}
Finally, with the toroidal SRR filled with only nitrogen vapor, we measured the quality factor of the copper resonator as it warmed from \SI{80}{K} to room temperature.  Since the toroidal geometry suppresses radiative losses, the $Q$ of the resonator is predominantly determined by the EM skin depth $\delta$ via
\begin{equation}
Q\approx \frac{r_0}{\delta},
\end{equation}
where $r_0$ is the radius of the bore of the resonator, \mbox{$\delta\approx\sqrt{2\rho/\left(\mu_0\omega\right)}$} for a good conductor, and $\rho$ is the resistivity of copper \cite{Hardy:1981, Bobowski:2013, Bobowski:2016}.  The crucial point for this measurement is that \mbox{$Q^{-2}\propto\rho$}.  In order to compare the quality factor data to published resistivity measurements, we define a dimensionless quantity that is proportional to the change in resistivity
\begin{equation}
\delta\rho_\mathrm{S}\equiv\frac{\rho(T)-\rho(T_0)}{\rho(T_\mathrm{f})-\rho(T_0)}=\frac{Q^{-2}(T)-Q^{-2}(T_0)}{Q^{-2}(T_\mathrm{f})-Q^{-2}(T_0)},
\end{equation} 
where, for our measurements, $T_0=\SI{80}{K}$ and $T_\mathrm{f}=\SI{222}{K}$.

\begin{figure}[t]
\centering{\includegraphics[width=0.95\columnwidth]{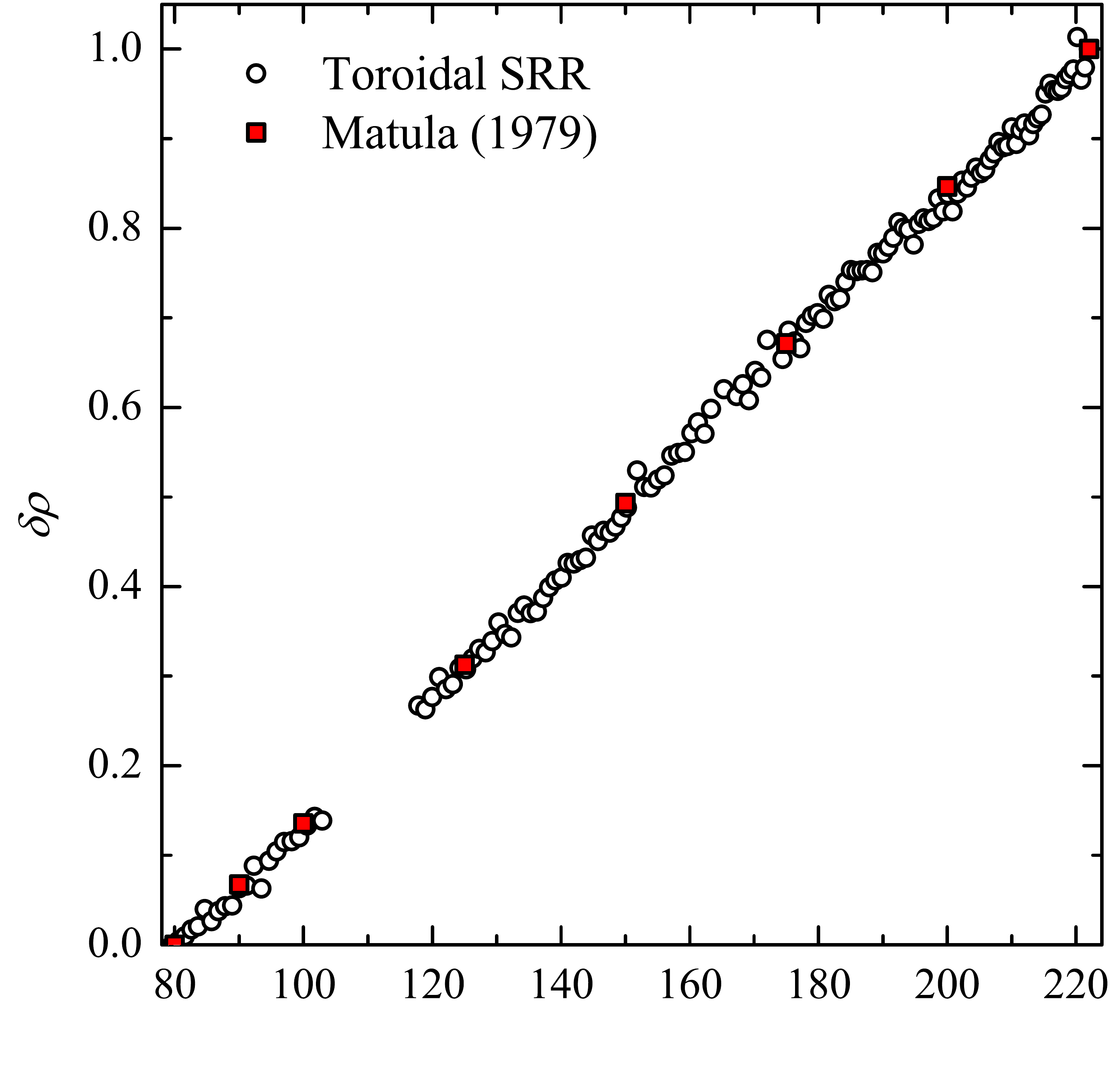}}
\caption{\label{fig:CuSigma}Scaled change in resistivity calculated from the quality factor measurements (white circles) compared to values determined from resistivity data published in \cite{Matula:1979} (red squares).}
\end{figure}

A typical room-temperature $Q$ of the copper resonator is 2000 (see Table~\ref{tab:fits}) whereas, at liquid nitrogen temperatures, the $Q$ exceeds 5000. The $\delta\rho_\mathrm{S}$ data determined from the quality factor measurements are shown as the white circles in Fig.~\ref{fig:CuSigma} and show a resistivity that varies linearly with temperature. The gap in the data between \SI{100} and \SI{120}{K} was caused by an unexpected interruption to the data acquisition program. 
The measured data are compared to $\delta\rho_\mathrm{S}$ calculated from published values of copper's resistivity (red squares), and the agreement is excellent \cite{Matula:1979}. 

The uncertainty in the $\delta\rho_\mathrm{S}$ measurements, attributed to the drift in the quality factor during the \SI{30}{\second} frequency sweeps, are largest at temperatures near $T_\mathrm{f}$.  The drift rate of $Q(T)$ was less than \SI{30}{\per\minute} which corresponds to $\Delta Q(T)\approx 15$.  Using propagation of errors, the uncertainty in $\delta\rho_\mathrm{S}$ in the limit that $T\to T_\mathrm{f}$ is given by
\begin{equation}
\Delta\left(\delta\rho_\mathrm{S}\right)\approx\frac{\sqrt{2}\Delta Q(T)/Q^3(T)}{Q^{-2}\left(T_\mathrm{f}\right)-Q^{-2}\left(T_0\right)}
\end{equation}
which evaluates to $\approx\SI{0.01}{}$ and is nearly equal to the size of the data points in Fig.~\ref{fig:CuSigma}.

\section{Conclusions}\label{sec:summary}
The toroidal SRR is a compact and high-$Q$ RF resonator that can be used to characterize the EM properties of materials.  The electric fields are concentrated within the gap of the resonator and the magnetic fields within its bore.  This configuration allows one to separately characterize the electric and magnetic properties of a material.  In this paper, we have demonstrated the use of a toroidal SRR to make high-precision measurements of the complex permittivity of methanol at \SI{185}{MHz} (Section~\ref{sec:meth}), the dielectric constant of an atmosphere of air (Section~\ref{sec:gas}), the dielectric constant of liquid nitrogen at \SI{890}{MHz} from \SIrange[range-units = single]{63}{77}{\kelvin} (Section~\ref{sec:LN2}), and the temperature dependence of the resistivity of copper from \SIrange[range-units = single]{80}{220}{\kelvin} (Section~\ref{sec:CuSigma}).

Split-ring resonators are easy to design and fabricate making them a convenient choice for precision characterization of EM material properties.  Furthermore, the measurements typically require only very standard pieces RF test equipment and the data analysis techniques are often relatively straightforward, requiring only algebraic manipulations and conventional least-squares minimizations.  The toroidal geometry is particularly useful as it allows for high-$Q$ measurements using a compact experimental testbed. 

We next plan to use the toroidal SRR to characterize the magnetic properties of a suspension of ultrasmall superparamagnetic iron oxide (USPIO) nanoparticles.  For example, with the USPIO suspension contained within the bore of the toroidal SRR, one could measure its complex permeability as a function of either the nanoparticle concentration or an applied DC magnetic field.

One obvious limitation of the toroidal SRR is that it is a fixed-frequency technique.  However, if the bore of the resonator is loaded with a soft ferrite material, such as Ni$_x$Zn$_{1-x}$Fe$_2$O$_4$, one could use an external magnetic field to tune the permeability of the ferrite and, therefore, the resonance frequency of the SRR.  If, at the same time, the gap of the resonator is filled with a dielectric material, one could in principle make high-resolution spectroscopic measurements of the complex permittivity of that material over a limited frequency range.  This technique would require the imaginary component of the ferrite's permeability to remain sufficiently small over the frequencies of interest.

\section*{Acknowledgment}

We gratefully acknowledge the support of Thomas Johnson who provided access to the Agilent N5241A VNA.

\ifCLASSOPTIONcaptionsoff
  \newpage
\fi



\bibliographystyle{IEEEtran}
%

%

\begin{IEEEbiography}[{\includegraphics[width=1in,height=1.25in,clip,keepaspectratio]{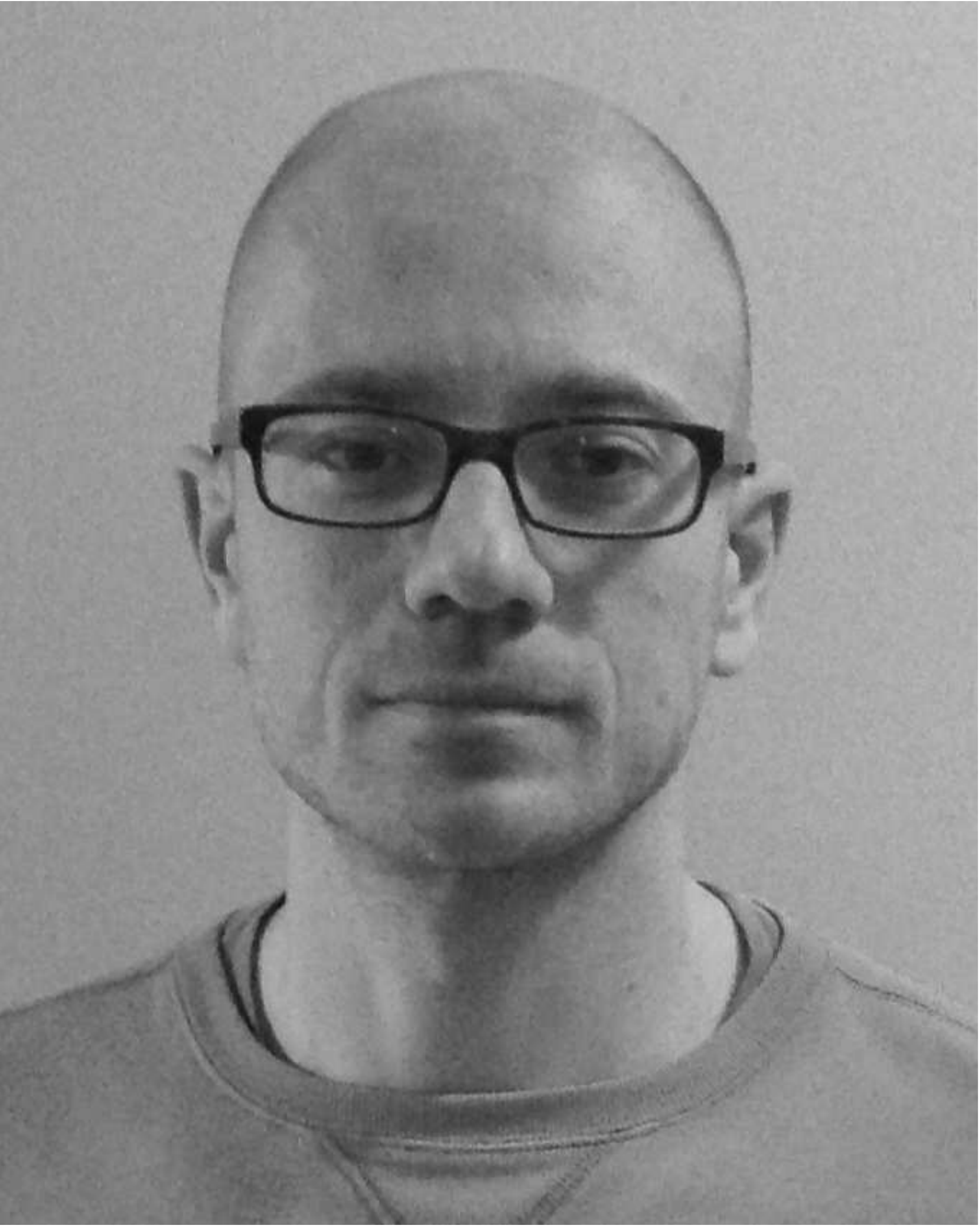}}]{Jake S. Bobowski}
  was born in Winnipeg, Canada on March 9, 1979.  He received a B.Sc. degree in physics from the University of Manitoba, Canada in 2001.  He was awarded M.Sc. and Ph.D. degrees in physics from the University of British Columbia, Canada in 2004 and 2010, respectively.  From 2011 to 2012, he was a postdoctoral fellow in the RF and Microwave Technology Research Laboratory in the Department of Electrical Engineering, and is now a senior instructor in physics, at the Okanagan campus of the University of British Columbia, Canada.  He is interested in developing custom microwave techniques to characterize the electromagnetic properties of a wide range of materials.
\end{IEEEbiography}

\begin{IEEEbiography}[{\includegraphics[width=1in,height=1.25in,clip,keepaspectratio]{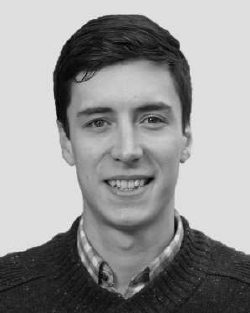}}]{Aaron P. Clements}
  was born in Nelson, New Zealand in 1993. He received the B.Sc. degree with honors in physics from the University of British Columbia, Kelowna, BC in 2016.
As a research assistant at UBC's Okanagan campus, he has worked on a range of research topics which include harnessing non-Newtonian fluid rheology in a novel pump design and non-contact detection of high-voltage hazards. For his honors project he used a toroidal split-ring resonator to implement an electron spin resonance experiment at \SI{1}{GHz}. He is currently employed as a research assistant in the School of Engineering at UBC's Okanagan campus, focused on resonant structure design for wireless power applications.
\end{IEEEbiography}





\end{document}